\documentclass[12pt]{article}
\usepackage{amsfonts}
\usepackage{amsmath}
\usepackage{amssymb}
\usepackage{graphicx}
\usepackage{color}
\usepackage[all, knot]{xy}
\usepackage{tikz}

\usepackage{ulem}

\usepackage{hyperref}
\usepackage[utf8]{inputenc}
\usepackage{epstopdf}
\usepackage[footnotesize]{caption}
\usepackage{amsthm}
\usepackage{enumitem}
\usepackage{mathrsfs}

\usepackage[margin=3cm]{geometry}
\def \be {\begin{equation}}
\def \ee {\end{equation}}
\def \bea {\begin{eqnarray}}
\def \eea {\end{eqnarray}}
\def \nn {\nonumber}

\def \rr {\raise.35ex\hbox{\small $\prime$}\kern-.17em{\mbox{\large $\imath$}}}

\def \dels {\partial\kern-.6em /\kern.1em}
\def \As {{A\kern-.5em / \kern.5em}}
\def \Ds {D\kern-.7em / \kern.5em}

\def \ks {k\kern-.5em /}
\def \ls {l\kern-.5em /}

\newcommand{\ci}[1]{}

\newcommand{\ba}{\begin{eqnarray}}
\newcommand{\ea}{\end{eqnarray}}
\newcommand{\bal}{\begin{align}}
\newcommand{\eal}{\end{align}}
\newcommand{\bay}[1]{\left(\begin{array}{#1}}
\newcommand{\eay}{\end{array}\right)}

\newcommand{\hide}[1]{}

\newlist{axioms}{enumerate}{2}
\setlist[axioms,1]{label=\textbf{A\arabic{axiomsi}.}, ref=A\arabic{axiomsi}}
\setlist[axioms,2]{label=\textbf{A\arabic{axiomsi}\rlap{\myEnumCounter{axiomsii}}.},%
                   ref=A\arabic{axiomsi}\myEnumCounter{axiomsii},%
                   align=parleft,%
                   leftmargin=0em,%
                   itemsep=1.4ex,%
                   before={\stepcounter{axiomsi}}}

  \usetikzlibrary{decorations.markings}

\begin{document}

\begin{titlepage}
\begin{center}

\textbf{\LARGE
Explore the Origin of\\ 
Spontaneous Symmetry Breaking from 
Adaptive Perturbation Method
\vskip.3cm
}
\vskip .5in
{\large
Chen-Te Ma$^{a,b,c,d,e,f}$ \footnote{e-mail address: yefgst@gmail.com},
Yiwen Pan$^{g}$ \footnote{e-mail address: panyw5@mail.sysu.edu.cn}, and 
Hui Zhang$^{c,e}$ \footnote{e-mail address: mr.zhanghui@m.scnu.edu.cn}
\\
\vskip 1mm
}
{\sl
$^a$
Department of Physics and Astronomy, Iowa State University, Ames, Iowa 50011, US. 
\\
$^b$
Asia Pacific Center for Theoretical Physics,\\
Pohang University of Science and Technology, 
Pohang 37673, Gyeongsangbuk-do, South Korea. 
\\
$^c$
Guangdong Provincial Key Laboratory of Nuclear Science,\\
 Institute of Quantum Matter,
South China Normal University, Guangzhou 510006, Guangdong, China. 
\\ 
$^d$ 
School of Physics and Telecommunication Engineering,\\
South China Normal University, Guangzhou 510006, Guangdong, China. 
\\ 
$^e$
Guangdong-Hong Kong Joint Laboratory of Quantum Matter,\\
 Southern Nuclear Science Computing Center, 
South China Normal University, 
Guangzhou 510006, Guangdong, China. 
\\
$^f$ 
The Laboratory for Quantum Gravity and Strings,\\
 Department of Mathematics and Applied Mathematics, 
University of Cape Town,
 Private Bag, Rondebosch 7700, South Africa.
 \\
 $^g$
 School of Physics, Sun Yat-Sen University, Guangzhou 510275, Guangdong, China.
}\\
\vskip 1mm
\vspace{40pt}
\end{center}

\newpage
\begin{abstract} 
Spontaneous symmetry breaking occurs when the underlying laws of a physical system are symmetric, but the vacuum state chosen by the system is not. 
The (3+1)d $\phi^4$ theory is relatively simple compared to other more complex theories, making it a good starting point for investigating the origin of non-trivial vacua. 
The adaptive perturbation method is a technique used to handle strongly coupled systems. 
The study of strongly correlated systems is useful in testing holography. 
It has been successful in strongly coupled QM and is being generalized to scalar field theory to analyze the system in the strong-coupling regime. 
The unperturbed Hamiltonian does not commute with the usual number operator. 
However, the quantized scalar field admits a plane-wave expansion when acting on the vacuum. 
While quantizing the scalar field theory, the field can be expanded into plane-wave modes, making the calculations more tractable. 
However, the Lorentz symmetry, which describes how physical laws remain the same under certain spacetime transformations, might not be manifest in this approach. 
The proposed elegant resummation of Feynman diagrams aims to restore the Lorentz symmetry in the calculations. 
The results obtained using this method are compared with numerical solutions for specific values of the coupling constant $\lambda = 1, 2, 4, 8, 16$. 
Finally, we find evidence for quantum triviality, where self-consistency of the theory in the UV requires $\lambda = 0$. 
This result implies that the $\phi^4$ theory alone does not experience SSB, and the $\langle \phi\rangle = 0$ phase is protected under the RG-flow by a boundary of Gaussian fixed-points.
\end{abstract}
\end{titlepage}

\section{Introduction}
\label{sec:1}
\noindent
The self-interaction of the Higgs field \cite{Goldstone:1961eq} is one way to generate mass through quantum correction, but it faces challenges due to ultraviolet (UV) divergences. 
These divergences arise in calculations when considering interactions at extremely short distances or high energies, which indicates that the theory might break down at such scales. 
The UV divergence problem is a longstanding issue in Quantum Field Theory (QFT). 
The evidence presented in Ref. \cite{Luscher:1987ay} suggests that in a four-dimensional scalar field theory, the self-interaction of a scalar field alone might not be enough to generate the spontaneous symmetry breaking (SSB) responsible for mass generation. 
This has implications for the Higgs mechanism, as it indicates that additional matter interactions might be necessary for the observed mass generation. 
In this letter, we discuss the analysis of quantum corrections in (3+1)d $\phi^4$ theory, particularly when the scalar mass vanishes ($m_p=0$) \cite{Coleman:1973jx}. 
The $\phi^4$ theory in dimensions higher than four is known to exhibit triviality \cite{Aizenman:1981du}, but in four dimensions, it remains an open problem.
\\

\noindent
We are interested in determining whether a non-trivial vacuum $\langle 0|\phi|0\rangle\neq 0$ arises solely from semi-classical corrections without the influence of a non-zero physical mass. 
However, the analysis of quantum corrections in a strongly coupled theory presents challenges, and existing methods may not be feasible. 
One of the challenges \cite{Coleman:1973jx} arises from the lack of a suitable computation method in a strongly coupled theory. 
The adaptive perturbation method \cite{Weinstein:2005kw,Weinstein:2005kx} has proven to be reliable in the context of Quantum Mechanics (QM) beyond the weak-coupling regime \cite{Ma:2019pxd,Ma:2020ipi,Ma:2020syr}. 
The central question we aim to address in this letter is: {\it Uncovering the Origin of SSB from Adaptive Perturbation Method?}
\\

\noindent 
Following Refs. \cite{Weinstein:2005kw,Weinstein:2005kx} instead, we introduce a functional parameter $\gamma(\vec{k})$ (an even function for $\vec{k}$ and non-negative) to the scalar field $\phi$ and its conjugate momentum ($\Pi$). 
The scalar field can be expanded in terms of creation ($a^{\dagger}$) and annihilation ($a$) operators as follows:  
  \bea
  &&
\phi(x)|0\rangle
\nn\\
&\equiv& \bigg(\int \frac{d^3\vec{k}}{(2\pi)^3}\frac{1}{\sqrt{2\gamma(\vec{k})}}(a_{\vec{k}}e^{ikx}+a^{\dagger}_{\vec{k}}e^{-ikx})
+\phi_0\bigg)\bigg|0\bigg\rangle,
\label{sm}
 \eea
 where $\vec{k}$ represents the momentum, $k\cdot x \equiv - E_{\vec k}x^0 + \vec k \cdot \vec x$, $E_{\vec k} \equiv \vec k^2 + m^2$, and $m$ is the mass. 
 The constant  $\phi_0 \in \mathbb{R}$ represents a condensation of the scalar field. 
 When evaluated in the vacuum state $|0\rangle$, without a loop correction, the expectation value of the scalar field is $\langle \phi\rangle = \phi_0$. 
 The conjugate momentum $\Pi(x)$ is defined as the time derivative of the scalar field 
 \bea
 \Pi(x)|0\rangle\equiv\partial_0\phi(x)|0\rangle. 
 \eea 
 When $\phi_0 = 0$, the two expansions of $\phi$ are related by the Bogoliubov transformation: $U(\theta)a_{\vec{k}}U^{-1}(\theta)\rightarrow a_{\vec{k}}$ and $U(\theta)|0\rangle\rightarrow|0\rangle$, where $U(\theta)$ is a unitary operator parameterized by $\theta$.  
\\

\noindent 
In the context of perturbation theory, the choice of the unperturbed Hamiltonian $H_0$ plays a crucial role in determining how the perturbative expansion is performed. 
The different decomposition does not affect the canonical variables but the convenience of the expansion. 
In conventional perturbation theory, $H_0$ is usually chosen to be the free part of the Hamiltonian, representing the system without any interactions. 
This choice simplifies the perturbative calculations and is often convenient in many physical situations. 
Different choices of $H_0$ can lead to different perturbative expansions, but they should ultimately yield the same physical predictions. 
The adaptive perturbation method \cite{Weinstein:2005kw,Weinstein:2005kx} made a non-conventional choice for $H_0$. 
To describe their approach briefly, they rewrote the total Hamiltonian in terms of creation and annihilation operators. 
Then, they specified $H_0$ as the sum of terms with an equal number of creation and annihilation operators. 
Ultimately, the validity and usefulness of a particular choice for $H_0$ depend on the specific physical system and the goals of the perturbative calculations. 
The key requirement is that the chosen $H_0$ should be tractable enough to allow for perturbative treatment and that the perturbative results are consistent with the physical reality of the system under consideration. 
\\

\noindent 
In this paper, we focus on determining $\gamma(\vec{k})$ and $\phi_0$ by minimizing the energy. 
The initial step lays the foundation for further exploration. 
Subsequently, we successfully derive the plane-wave solution when the fields act on the vacuum state, as defined in Eq. \eqref{sm}. 
This paves the way for a deeper understanding of the system under consideration. 
It is important to mention that while the adaptive perturbation method has been well-established in its Hamiltonian formulation \cite{Schwinger:1951ex}, we also explore the possibility of developing a Lagrangian description. 
We expect that a Lagrangian perspective could complement our findings and lead to even more comprehensive insights. 
One goal is to investigate Lorentz symmetry in the adaptive perturbation method. 
To achieve this, we have employed a novel approach of resuming Feynman diagrams, organized by powers of both $\phi_0$ and $\gamma(\vec{k})$. 
This extension of the method allows us to incorporate Lorentz symmetry effectively, opening up new avenues of investigation. 
Furthermore, we utilize the diagrammatic method to compute correlation functions, enabling us to study the system even under conditions of significant coupling. 
The successful matching of these correlation functions with lattice simulation results demonstrates the accuracy and robustness of our approach. 
One of the key insights we obtain from our analysis, using the renormalization group (RG) flow \cite{Gell-Mann:1954yli}, is that the self-interaction of a scalar possible not be the primary origin of SSB. 
Instead, we find evidence suggesting that interactions with other matters play a crucial role in the process of SSB. 
This finding opens up intriguing possibilities for future research in this field. 

\section{Dynamics}
\label{sec:2} 
\noindent
The Hamiltonian for $\lambda\phi^4$ theory is given by
 \bea
 H&=&\int d^3x\ \bigg(\frac{1}{2}\Pi^2(x)+\frac{1}{2}\partial_j\phi(x)\partial_j\phi(x)+\frac{m^2}{2}\phi^2(x)
 \nn\\
 &&+\frac{\lambda}{4!}\phi^4(x)\bigg). 
 \eea
The index $j=1, 2, 3$ denotes the space dimensions, while the parameters $m$ and $\lambda$ are the bare mass and the (positive) coupling constant, respectively.
\\

\noindent 
The parameters $\gamma(\vec k)$ and $\phi_0$ are not arbitrary. 
More precisely, we fix them by minimizing the vacuum energy. 
The direct computation shows that vacuum energy is
\bea
&&
\langle 0| H|0\rangle
\nn\\
&=&
V\bigg\lbrack
\frac{m^2}{2}\phi_0^2
+\frac{\lambda}{4!}\phi_0^4
+\int\frac{d^3\vec{k}}{(2\pi)^3}\ \bigg(\frac{\gamma(\vec{k})}{4}
+\frac{\vec{k}\cdot\vec{k}+m^2}{4\gamma(\vec{k})}
\bigg)
\nn\\
&&
+\frac{\lambda}{8}\phi_0^2\int\frac{d^3\vec{k}}{(2\pi)^3}\ \frac{1}{\gamma(\vec{k})}
+\frac{\lambda}{32}\bigg(\int\frac{d^3\vec{k}}{(2\pi)^3}\ \frac{1}{\gamma(\vec{k})}\bigg)^2
\bigg\rbrack, 
\eea 
where $V$ is a spatial volume. 
\\
 
\noindent 
Requiring $\gamma(\vec k)$ to extremize $\langle H\rangle$ imposes the first saddle point equation,
\begin{equation}
    \label{1st-saddle}
    \gamma^2(\vec k)-( \vec k^2+m^2 ) -\frac{\lambda \phi _0^2}{2}-\frac{\lambda}{4}\int \frac{d^3 p}{(2\pi)^3} \frac{1}{\gamma(\vec p)} =0 \ , \qquad
    \forall \vec k\ .
\end{equation}
The arbitrariness in $\vec k$ allows the decomposition of this condition into two:
\bea
\gamma^2 (\vec k)  
&=& \vec{k}\cdot\vec{k}+\gamma^2(0), 
\nn\\ 
\gamma^2(0)&=&m^2+\frac{\lambda}{2}\phi_0^2
+\frac{\lambda}{4}\int\frac{d^3\vec{k}}{(2\pi)^3}\ \frac{1}{\gamma(\vec{k})} \ . 
\label{vq}
\eea
Requiring $\phi_0$ to extremize $\langle H\rangle$ imposes the second saddle point equation,
\bea
\label{2nd-saddle}
\phi_0\bigg(m^2+\frac{\lambda}{3!}\phi_0^2
+\frac{\lambda}{4}\int\frac{d^3\vec{k}}{(2\pi)^3}\ \frac{1}{\gamma(\vec{k})}\bigg)=0 \ .
\eea
The Eq. \eqref{2nd-saddle} admits two solutions, one with $\phi_0 = 0$, and the other generically non-zero determined by $m$ and $\lambda$. 
In what follows, we will frequently treat cases with $\phi_0 = 0$ and $\phi_0 \ne 0$ separately. 
Note that when $m^2 \ge0$, the expression in the parenthesis is non-zero. 
Therefore, one must pick the $\phi_0 = 0$ solution.
\\

\noindent 
The commutator of $H_0$ and $a^{\dagger}_{\vec{p}}$ contains terms with different numbers of $a$ and $a^\dagger$,
 \bea
 &&
 \lbrack H_0, a_{\vec{p}}^{\dagger}\rbrack
 \nn\\
 &=&
 \frac{\gamma_{\vec{p}}}{2}a_{\vec{p}}^{\dagger}
 +\frac{\vec{p}\cdot\vec{p}+m^2}{2\gamma_{\vec{p}}}a^{\dagger}_{\vec{p}}
 +\frac{\lambda\phi_0^2}{4\gamma_{\vec{p}}}a_{\vec{p}}^{\dagger}
 \nn\\
 &&
 +\frac{\lambda}{4!}\int\frac{d^3\vec{k}_1}{(2\pi)^3}\int\frac{d^3\vec{k}_2}{(2\pi)^3}\int\frac{d^3\vec{k}_3}{(2\pi)^3}\ 
 \frac{1}{\sqrt{\gamma_{\vec{k}_1}\gamma_{\vec{k}_2}\gamma_{\vec{k}_3}\gamma_{\vec{p}}}}
 \nn\\
 &&\times
 (2\pi)^3\delta^{(3)}(\vec{k}_1+\vec{k}_2+\vec{k}_3+\vec{p})
 \nn\\
 &&
\times
 (a^{\dagger}_{-\vec{k}_1}a^{\dagger}_{-\vec{k}_2}a_{\vec{k}_3}
 +a_{\vec{k}_1}a^{\dagger}_{-\vec{k}_2}a^{\dagger}_{-\vec{k}_3}
 +a^{\dagger}_{-\vec{k}_1}a_{\vec{k}_2}a^{\dagger}_{-\vec{k}_3}).
 \nn\\
 \eea
 When the commutator acts on the vacuum, the time evolution of fields is the same as in quantizing a non-interacting theory 
 \bea
 &&
  \lbrack H_0, a_{\vec{p}}^{\dagger}\rbrack|0\rangle
 =\gamma_{\vec{p}}a^{\dagger}_{\vec{p}}|0\rangle 
 \nn\\
 &&\Rightarrow
 \qquad
 e^{i H_0 t} a_{\vec p}^\dagger e^{- i H_0t} |0\rangle = e^{i \gamma_{\vec p}t} a_{\vec p}^\dagger |0\rangle\ .
 \eea
 This computation uses Eq. \eqref{vq}.
Hence we show that Eq. \eqref{sm} quantizes the interacting sector successfully. 
However, general operators acting on the vacuum, such as $\phi^2|0\rangle$, do not have such simple plane-wave expansion (there is no such issue in QM). 
Therefore, Lorentz symmetry is not manifest in each perturbation order. 
The restoration of Lorentz symmetry happens when taking into account all orders.

\section{Adaptive Perturbation Method from Resummation} 
\label{sec:3}
\begin{figure}
\begin{center}
\includegraphics[width=0.5\textwidth]{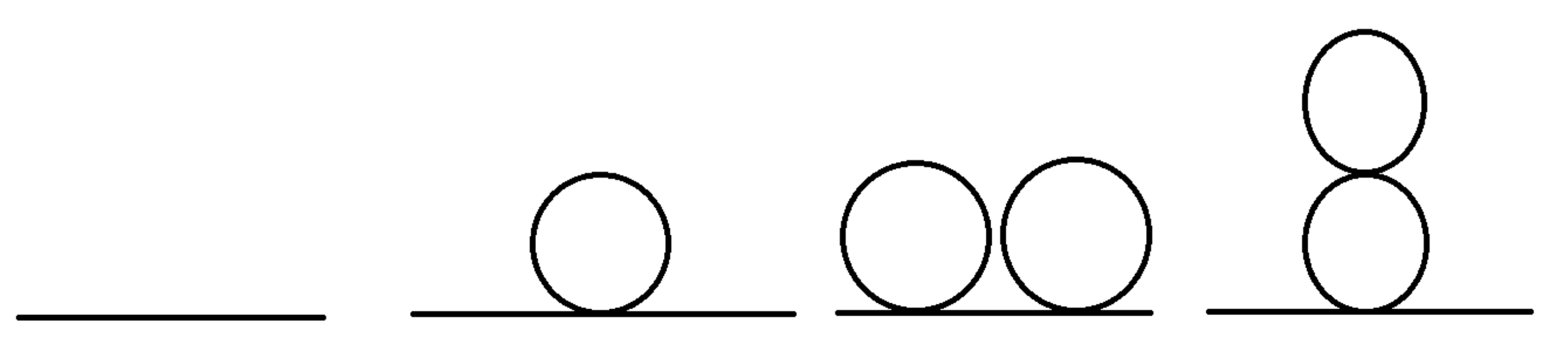}
\end{center}
\caption{This figure shows the summation of bubble diagrams for the two-point correlator with $\phi_0=0$ up to two loops in $\lambda\phi^4$ theory.} 
\label{sb}
\end{figure}

\noindent
Now we relate the canonical calculation to the Feynman diagrams at $t=0$. 
We start by considering $\phi_0 = 0$. 
In such a case, $\gamma^2(0)$ in Eq. \eqref{vq} can be solved recursively. 
Substituting the result into the two-point function, one obtains a summation of all bubble diagrams from a tree propagator. 
In Fig. \ref{sb}, we list the relevant diagrams up to two loops. 
The second and third diagrams arise as in the standard geometric series expansion. 
The last one is beyond the expansion. 
Note that although the recursion does not capture all the diagrams for the two-point function, the only missing one (up to two loops) is the sunset diagram with three internal propagators. 
\\

\noindent
We can reproduce the result upon replacing $m^2$ with the effective mass $\gamma^2(0)$, and compute $\langle \phi(x) \phi(y)\rangle$ directly from Eq. \eqref{sm} and the commutation relations. 
It is sufficient to analyze only the tree and sunset diagrams (up to two-vertex diagrams) for evaluating the connected two-point function. 
Hence the adaptive perturbation method should simplify the computation. 
\\

\begin{figure}
\begin{center}
\includegraphics[width=1.\textwidth]{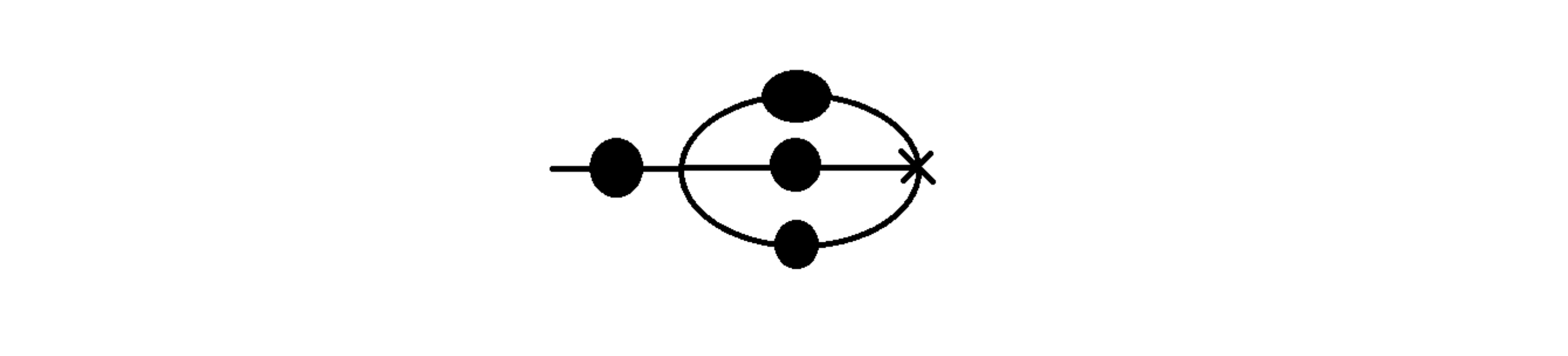}
\end{center}
\caption{The black dot in the diagram indicates the resummation by $\gamma^2(0)$. }
\label{1pt}
\end{figure}

\noindent 
Now we turn to the case with $\phi_0 \ne 0$. The Eq. \eqref{2nd-saddle} then implies
\begin{equation}
    \phi_0^2=-\frac{6m^2}{\lambda}-\frac{3}{2}\int\frac{d^3k}{(2\pi)^3} \frac{1}{\sqrt{2 \gamma(\vec k)}} \ .
    \end{equation}
With $\phi_0$ fixed by this equation, we compute the condensation $\langle \phi\rangle$ up to $\lambda^2$ order using Eq. \eqref{sm}, and we obtain
\bea
&&
\langle\phi\rangle - \phi_0
\nn\\
&\sim&
\phi_0\frac{\lambda^2}{4!\gamma^2(0)}\int\frac{d^3\vec{k}_1}{(2\pi)^3}\frac{1}{\gamma(\vec{k}_1)}
\nn\\
&&\times
\int\frac{d^3\vec{k}_2}{(2\pi)^3}\frac{1}{\gamma(\vec{k}_2)\gamma(\vec{k}_1+\vec{k}_2)
\big(\gamma(\vec{k}_1)+\gamma(\vec{k}_2)+\gamma(\vec{k}_1+\vec{k}_2)\big)
}. 
\eea 
Note that the VEV $\langle \phi\rangle \propto \phi_0$. Alternatively, the right-hand side corresponds to a single diagram shown in Fig. \ref{1pt}, where one applies the standard Feynman rule with $m^2$ replaced by the effective mass $\gamma^2(0)$. 
\\

\begin{figure}
\begin{center}
\includegraphics[width=1.\textwidth]{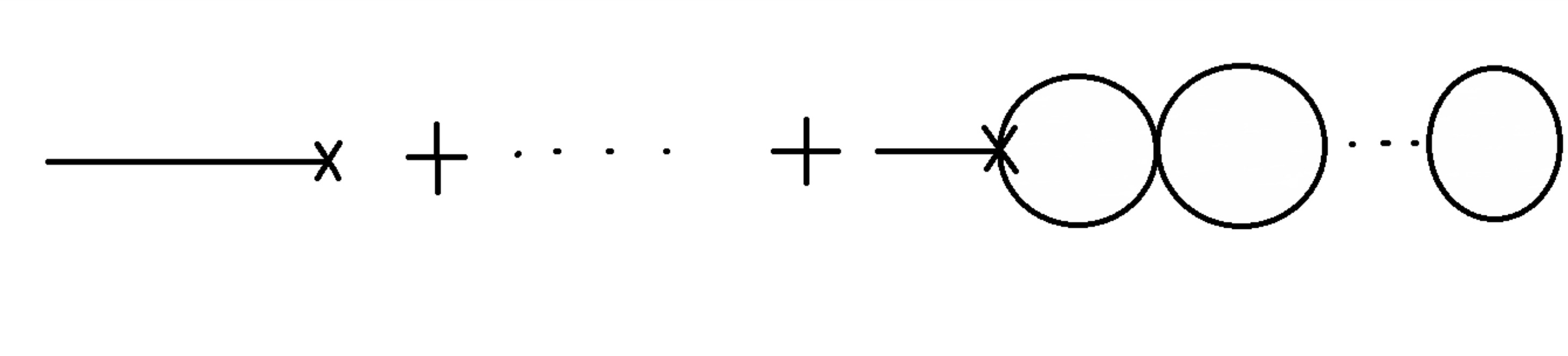}
\end{center}
\caption{This figure shows the bubble resummation for $\phi_0$. }
\label{rs1}
\end{figure}

\noindent
Although the above result is not manifestly Lorentz invariant, we expect that it is possible to restore the symmetry by hand from the following effective Lagrangian (here focuses on the case with $\phi_0 = 0$)
\bea
\mathcal{L}_0&=&-\frac{1}{2}(\partial_{\mu}\phi\partial^{\mu}\phi +\gamma^2(0)\phi^2), \ 
\nn\\
\mathcal{L}_I&=&\frac{\lambda}{4}\bigg(\int \frac{d^4q_E}{(2\pi)^4}\ \frac{1}{q_E^2+\gamma^2(0)}\bigg)\phi^2
+\frac{\lambda}{4!}\phi^4, 
\eea
where $\mathcal{L}_0$ and $\mathcal{L}_I$ are the unperturbed and perturbed parts, respectively. 
We label the Euclidean spacetime index as $\mu=0, 1, 2, 3$. 
The $q_E$ is the momentum in the Euclidean signature.
The unperturbed Hamiltonian $H_0$ is identical to the usual one but with the mass parameter given by $\gamma^2(0)$. 
Therefore, the usual Feynman rules apply with proper parameters. 
The number of relevant diagrams reduces significantly using the Lorentz invariant approach.
When performing perturbation computations using traditional methods, it is necessary to add up the contributions of all bubble diagrams shown in Fig. \ref{rs1} for $\langle \phi\rangle$ (include at least various one-loop and two-loop diagrams). 
In comparison, our proposal requires only one at two-loop, shown in Fig. \ref{1pt}. 
In this regard, our approach is an efficient resummation prescription of the Feynman diagrams.
   
\section{Perturbation vs. Lattice}
\label{sec:4}
\noindent 
We adopt the adaptive perturbation method to obtain an analytical solution and compare it with the lattice simulation. 
To compare with lattice simulation, we replace the propagator with the lattice propagator as in the following
\bea
&&
\int\frac{d^4k_E}{(2\pi)^4}\ \frac{1}{k_E^2+\gamma^2(0)}
\nn\\
&\rightarrow& 
\int\frac{d^4k_E}{(2\pi)^4}\ \frac{1}{\gamma^2(0)+\sum_{\mu=1}^4(2-2\cos(k_{E, \mu}))}.  
\eea
On the lattice, we set the lattice spacing $a$ to be 1. 
\\

\begin{figure}
\begin{center}
\includegraphics[width=0.32\textwidth]{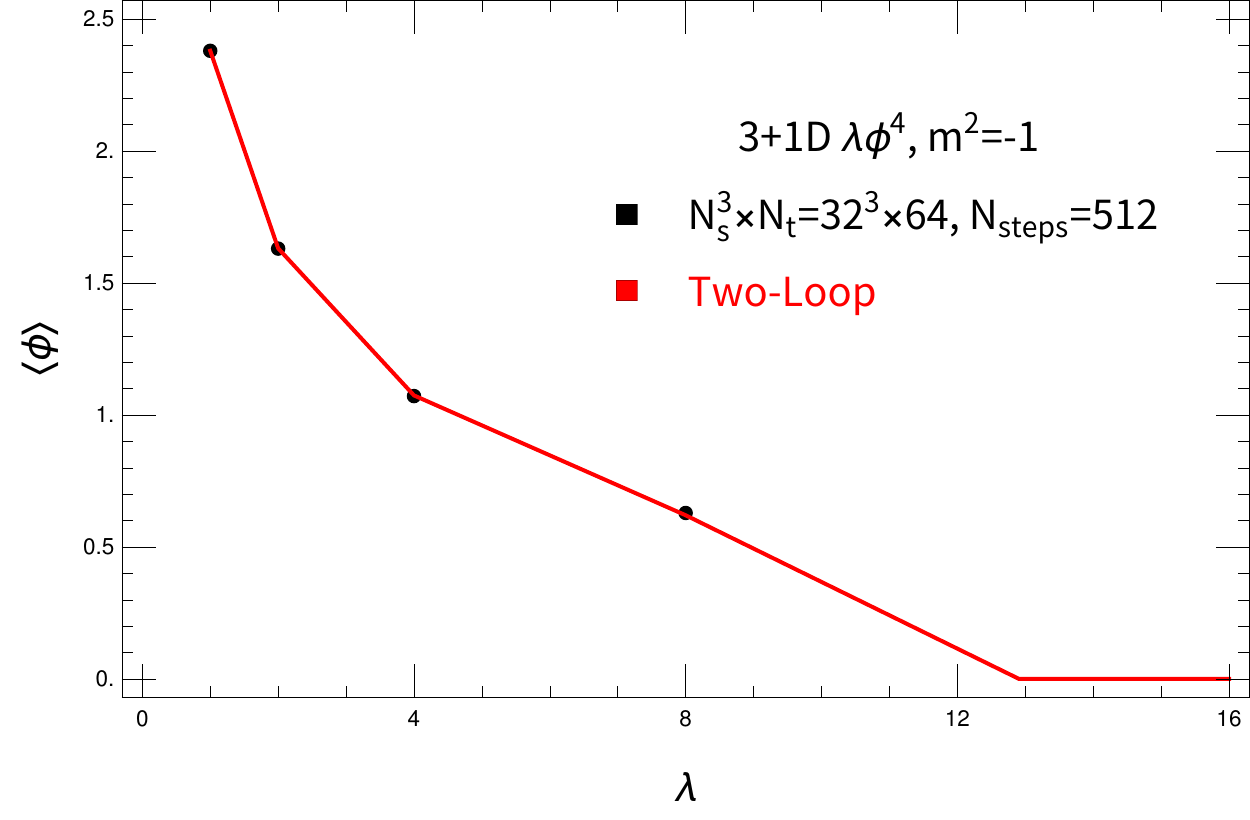}
\includegraphics[width=0.32\textwidth]{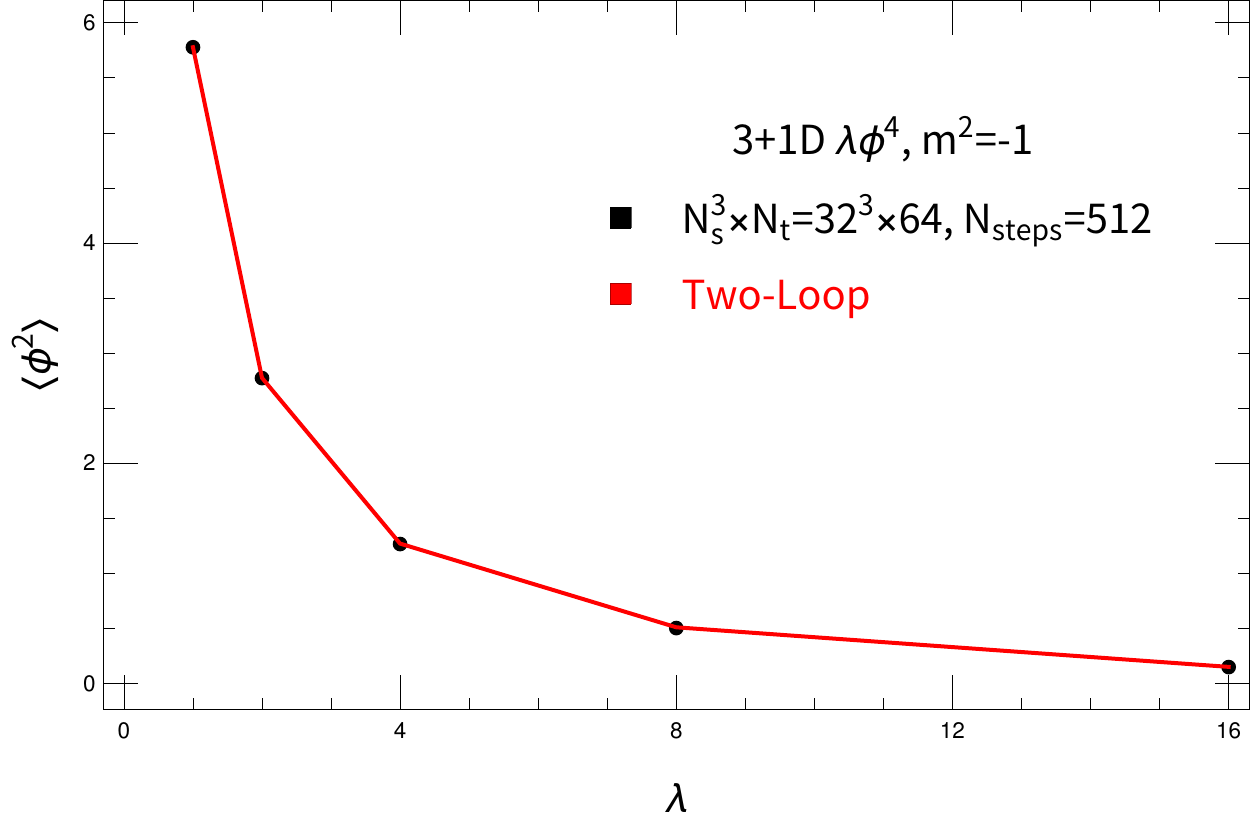}
\includegraphics[width=0.32\textwidth]{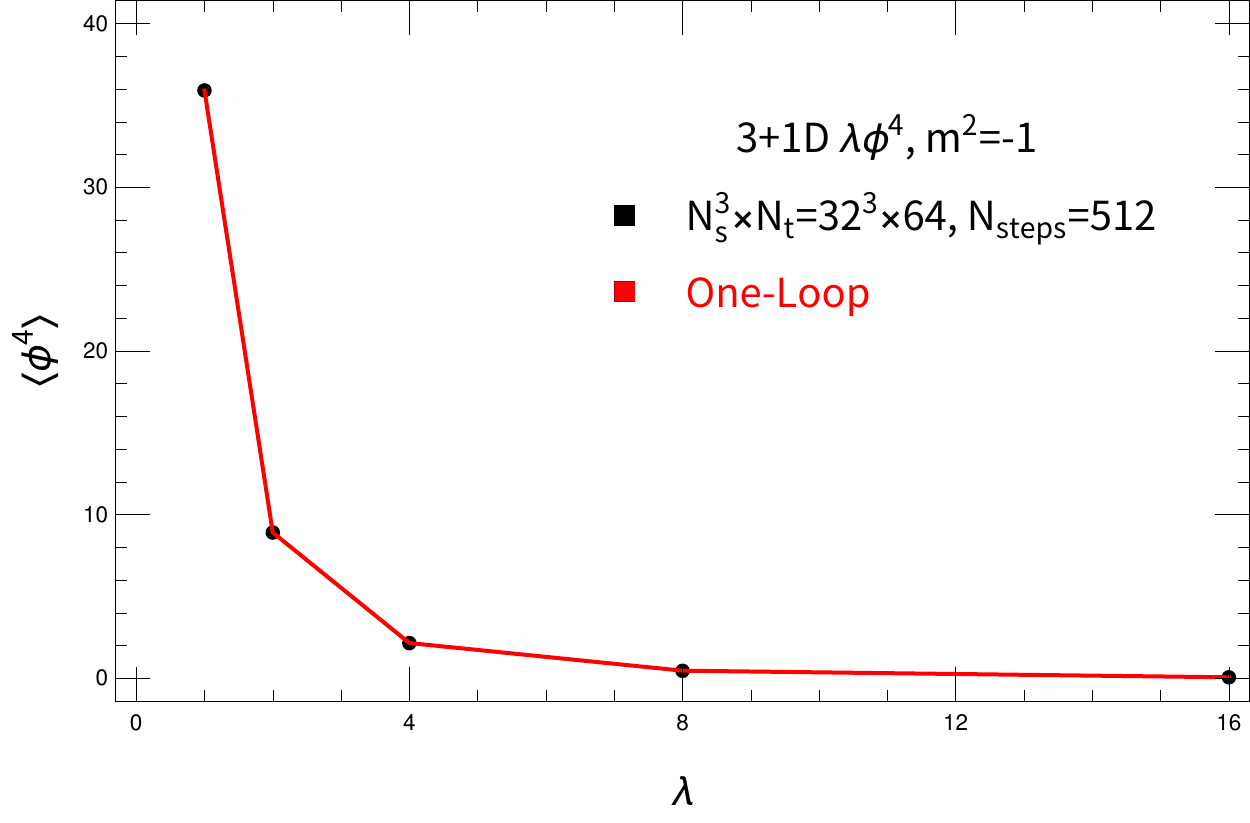}
\end{center}
\caption{We compare the perturbation result to the Hybrid Monte Carlo simulation in correlation functions. 
In the lattice simulation, we consider $\langle|\phi|\rangle$ rather than $\langle\phi\rangle$ for $\phi_0\neq 0$. 
The number of measurements is $2^{7}$ sweeps with thermalization $2^8$ sweeps and measure intervals $2^7$ sweeps. 
The error bars are less than $1\%$. 
The $N_{\mathrm{steps}}$ is the number of molecular dynamics steps. 
}
\label{pl}
\end{figure}

\begin{figure}
\begin{center}
\includegraphics[width=0.5\textwidth]{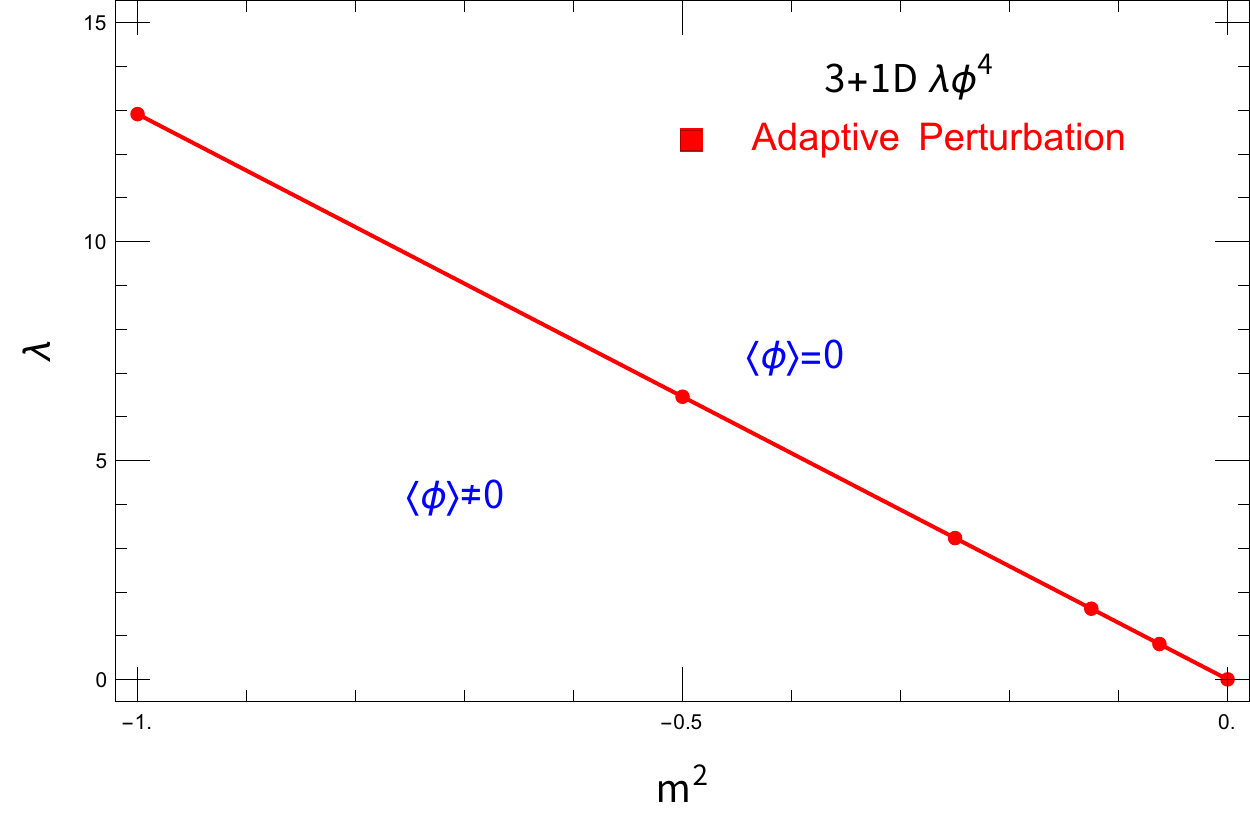}
\end{center}
\caption{The phase diagram of the critical line ($\gamma^2(0)=0$) separating by the symmetric and SSB cases. 
We show the perturbation result in the red dots and connect them as a straight line. 
This diagram illustrates the boundary between the two phases. 
}
\label{Transition}
\end{figure}

\noindent 
Now we use the solution of the saddle point:
\bea
&&
\Gamma^2(k_E)
\nn\\
&=&m^2+\frac{\lambda}{2}\phi_0^2
+\frac{\lambda}{2}\int\frac{d^4q_{E}}{(2\pi)^4}\ \frac{1}{q_{E}^2+\Gamma^2(q_E)}
\nn\\
&&
-\frac{\lambda^2\phi_0^2}{2}
\nn\\
&&\times
\int\frac{d^4q_E}{(2\pi)^4}\ \frac{1}{q_E^2+\Gamma^2(q_E)}\frac{1}{(k_E-q_E)^2+\Gamma^2(k_E-q_E)};
\nn\\
\eea
\bea
\phi_0\bigg(
 m^2+\frac{\lambda}{3!}\phi_0^2+\frac{\lambda}{2}\int\frac{d^4k_E}{(2\pi)^4}\ \frac{1}{k_E^2+\Gamma^2(k_E)}\bigg)=0, 
 \eea
 to reduce the number of diagrams (all one-loop diagrams are resumed). 
 We do the perturbation around the saddle point, and it matches the lattice simulation in Fig. \ref{pl}. 
Note that on a lattice, tunneling between different vacua cannot occur. 
Therefore, we simulate $\langle|\phi|\rangle$ and compare it to our result when $\phi_0\neq 0$. 
We simulate the lattice configurations from two vacuums. 
For $\phi_0 = 0$, one can only consider $\langle\phi^2\rangle$,$\langle\phi^4\rangle$ and compare them with lattice results. 
If the phase transition is continuous, the $\gamma^2(0)$ should be continuous across the two phases ($\langle \phi\rangle=0$ and $\langle\phi\rangle\neq0$). 
Said differently, if one solves the $\gamma^2(0)$ using the saddle point equations (Eqs. \eqref{1st-saddle}, \eqref{2nd-saddle}) with $\phi_0 \ne 0$, and then sends $\phi_0 \to 0$, continuity then implies that $\gamma^2(0)$ in this limit should equal the solution to Eqs. (\ref{1st-saddle}), (\ref{2nd-saddle}) with $\phi_0 = 0$. 
Such analysis leads to
\bea
m^2=-\frac{\lambda}{2}\int\frac{d^4k_E}{(2\pi)^4}\ \frac{1}{\sum_{\mu=1}^4(2-2\cos(k_{E, \mu}))}. 
\label{cp}
\eea
Note that when $m^2\ge 0$, the solution $\phi_0^2$ is non-positive (unphysical). 
Therefore, the physical solution can only be $\phi_0=0$. 
We show the phase diagram for $m^2\le 0$  in Fig. \ref{Transition}. 
Because $\phi$ is proportional to $\phi_0$, the analysis of the phase separation not just at the tree level but also loop effects.

\section{RG Flow}
\label{sec:5}
\begin{figure}
\begin{center}
    \includegraphics[width=0.49\textwidth]{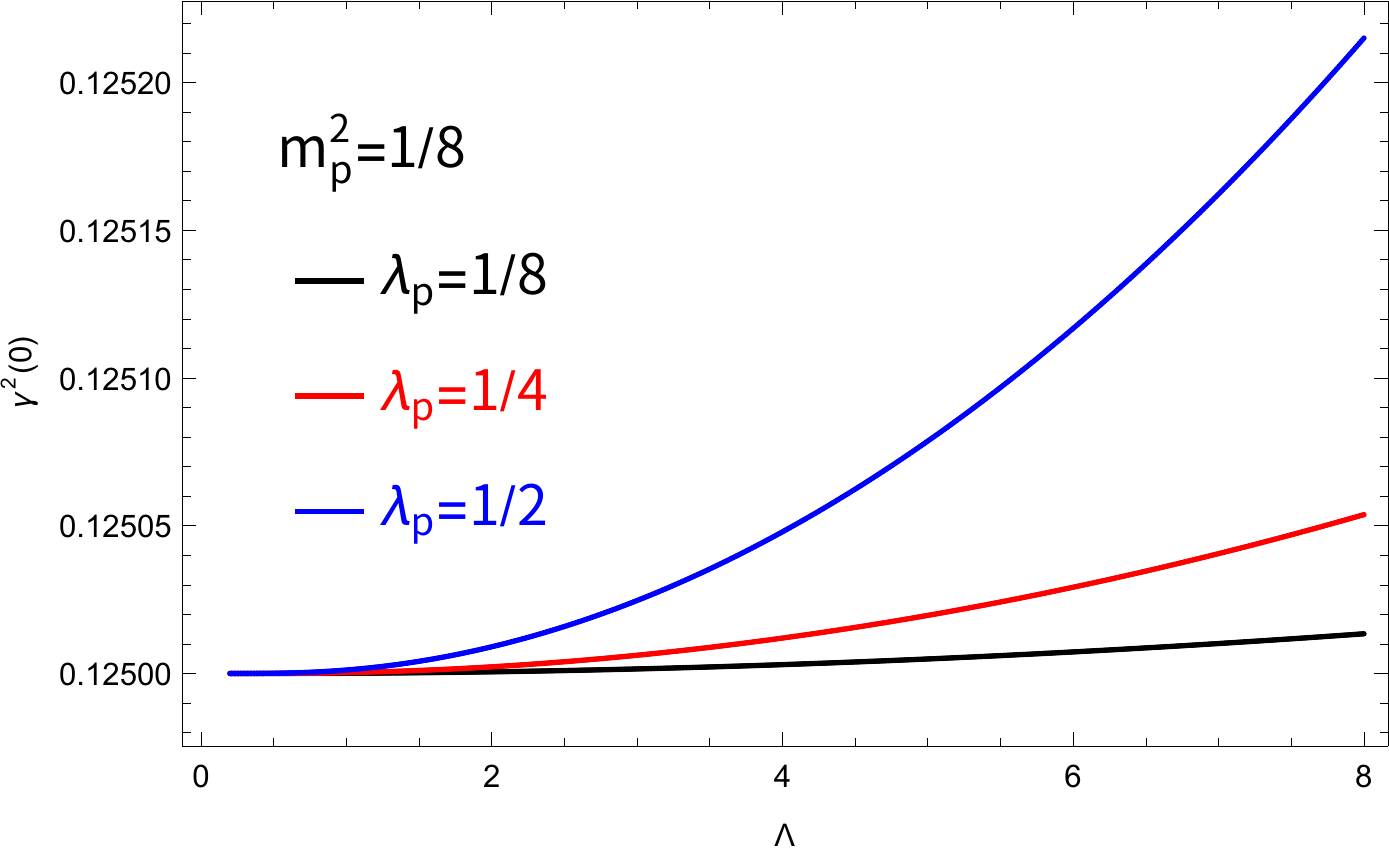}
    \includegraphics[width=0.49\textwidth]{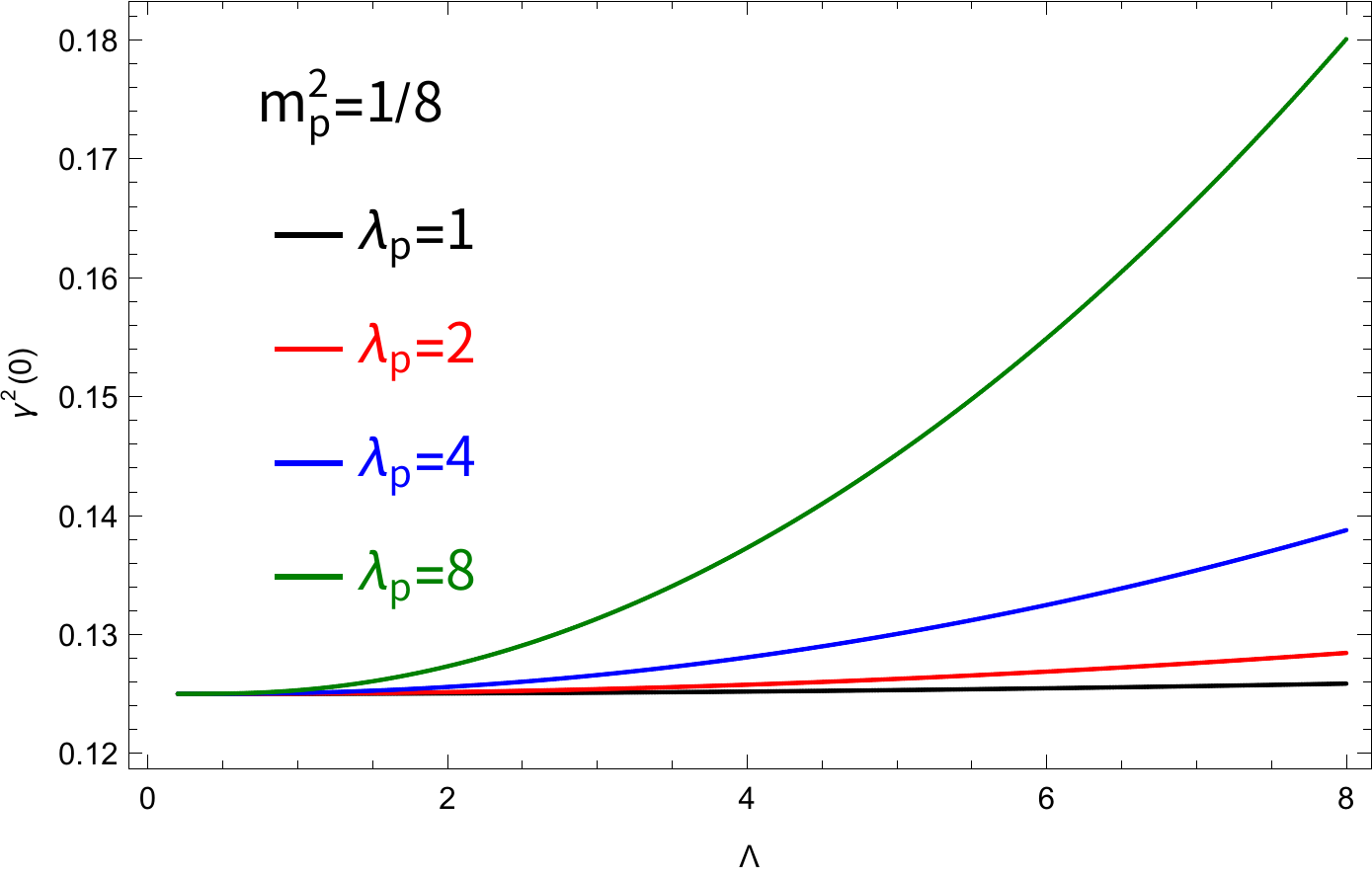}
    \includegraphics[width=0.49\textwidth]{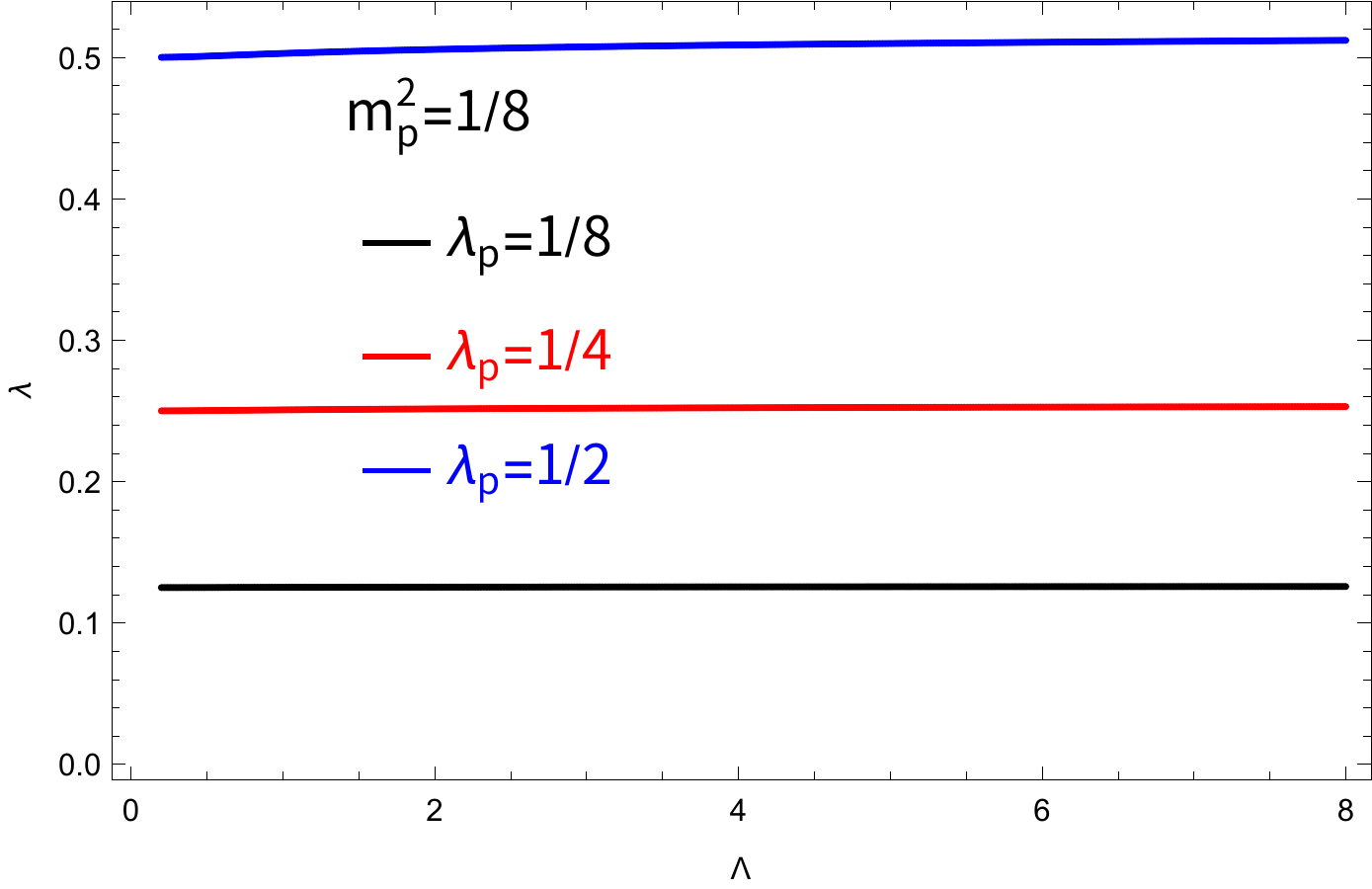}
    \includegraphics[width=0.49\textwidth]{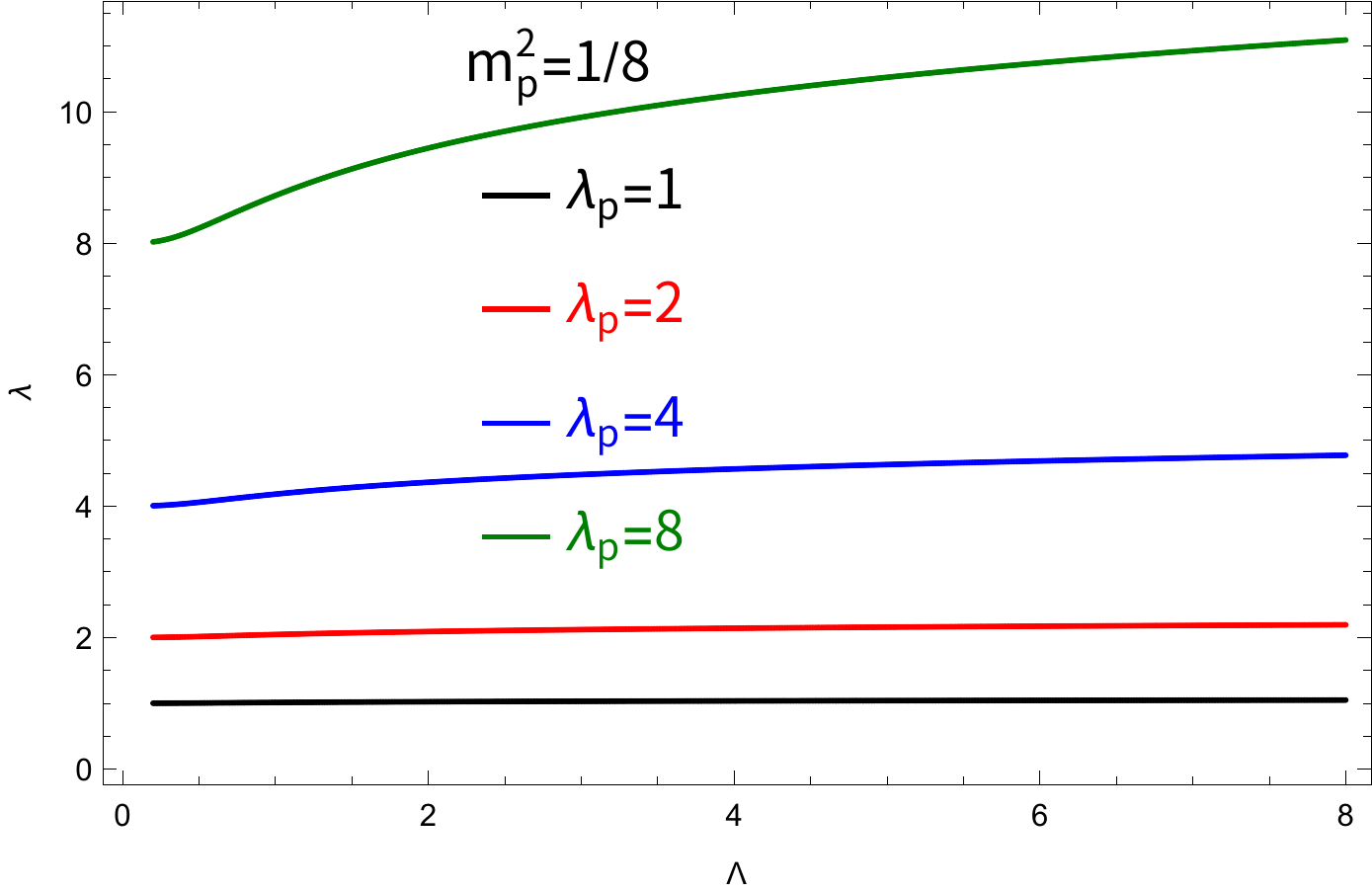}
    \includegraphics[width=0.49\textwidth]{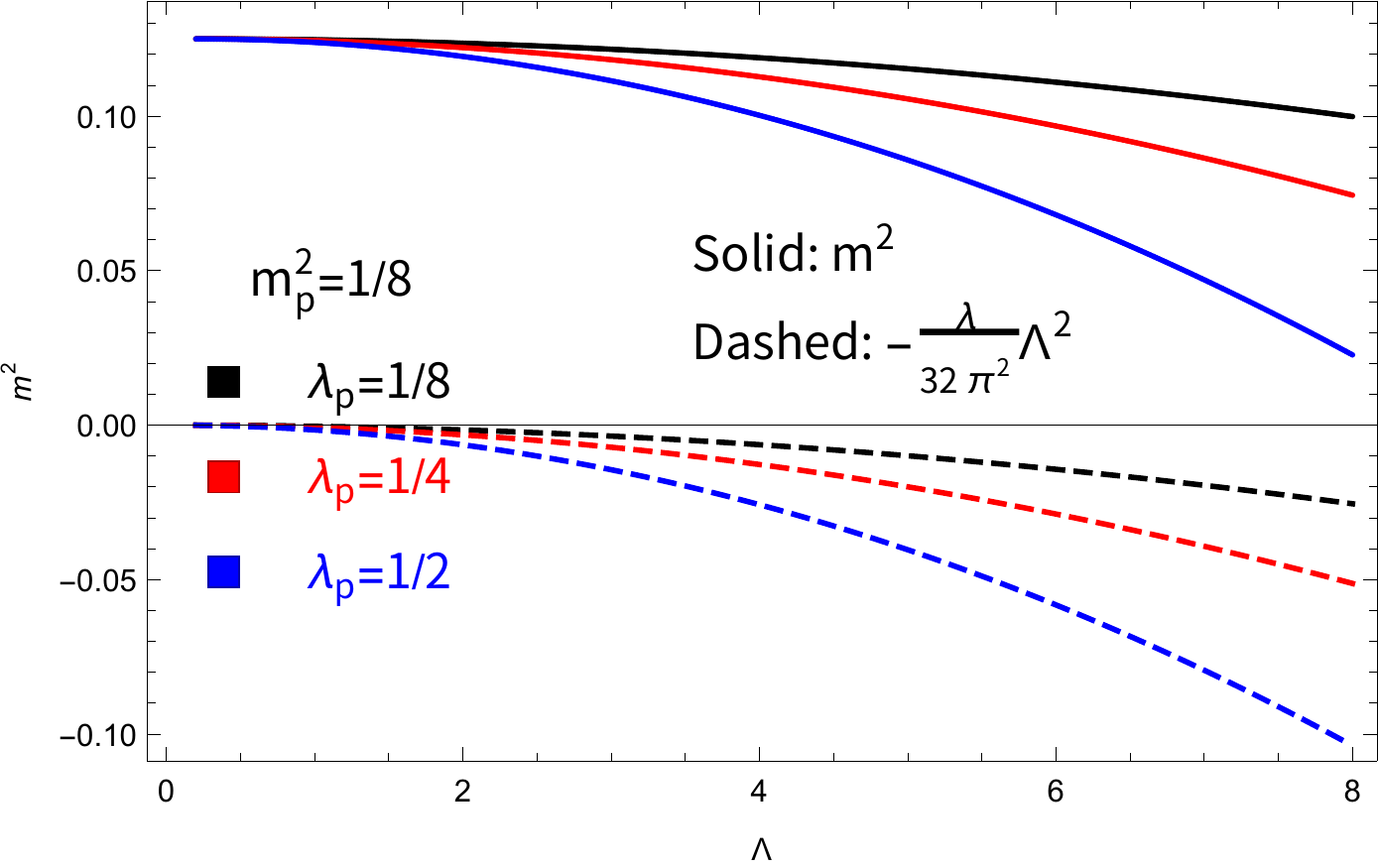}
    \includegraphics[width=0.49\textwidth]{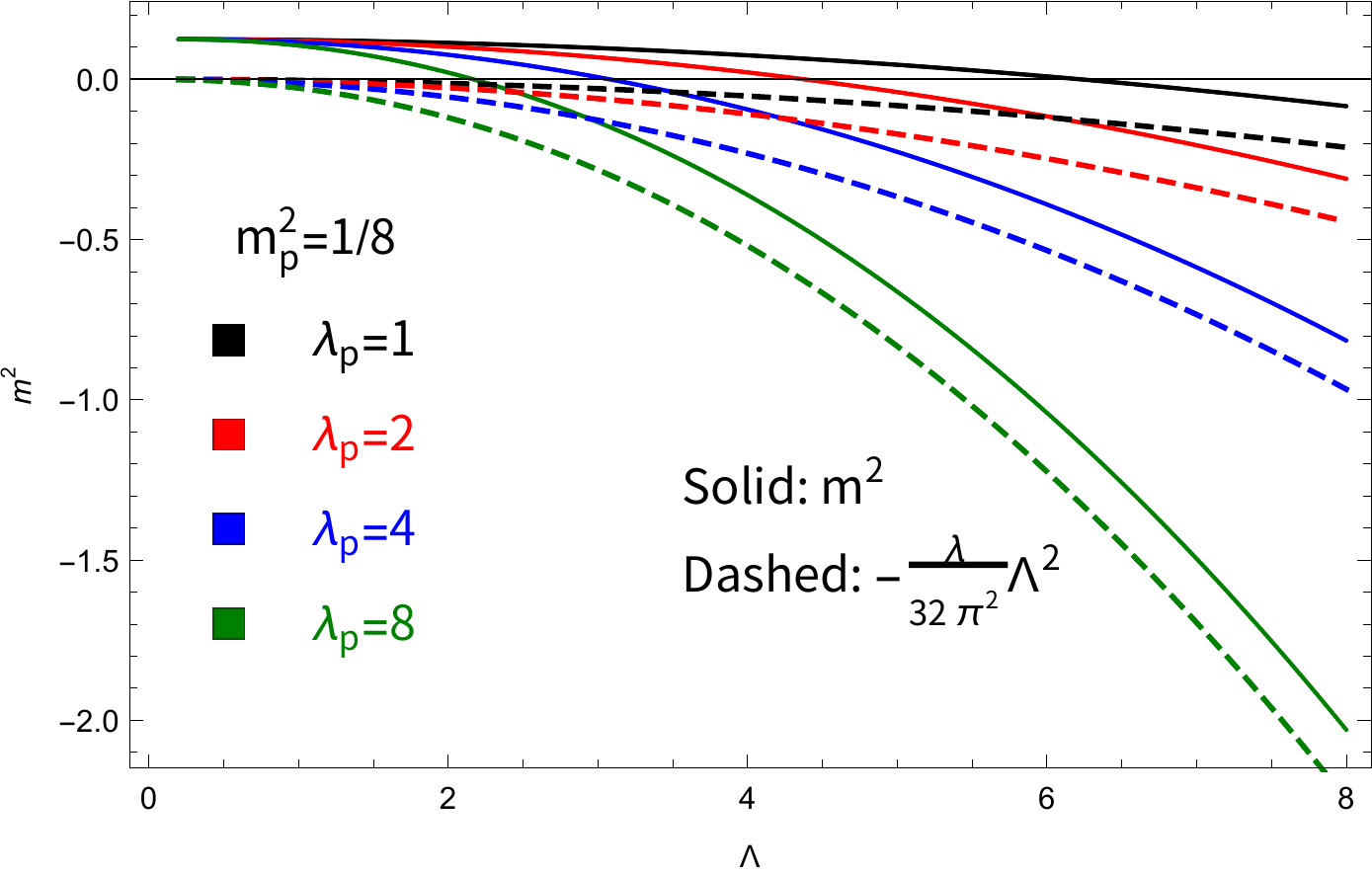}
    \includegraphics[width=0.49\textwidth]{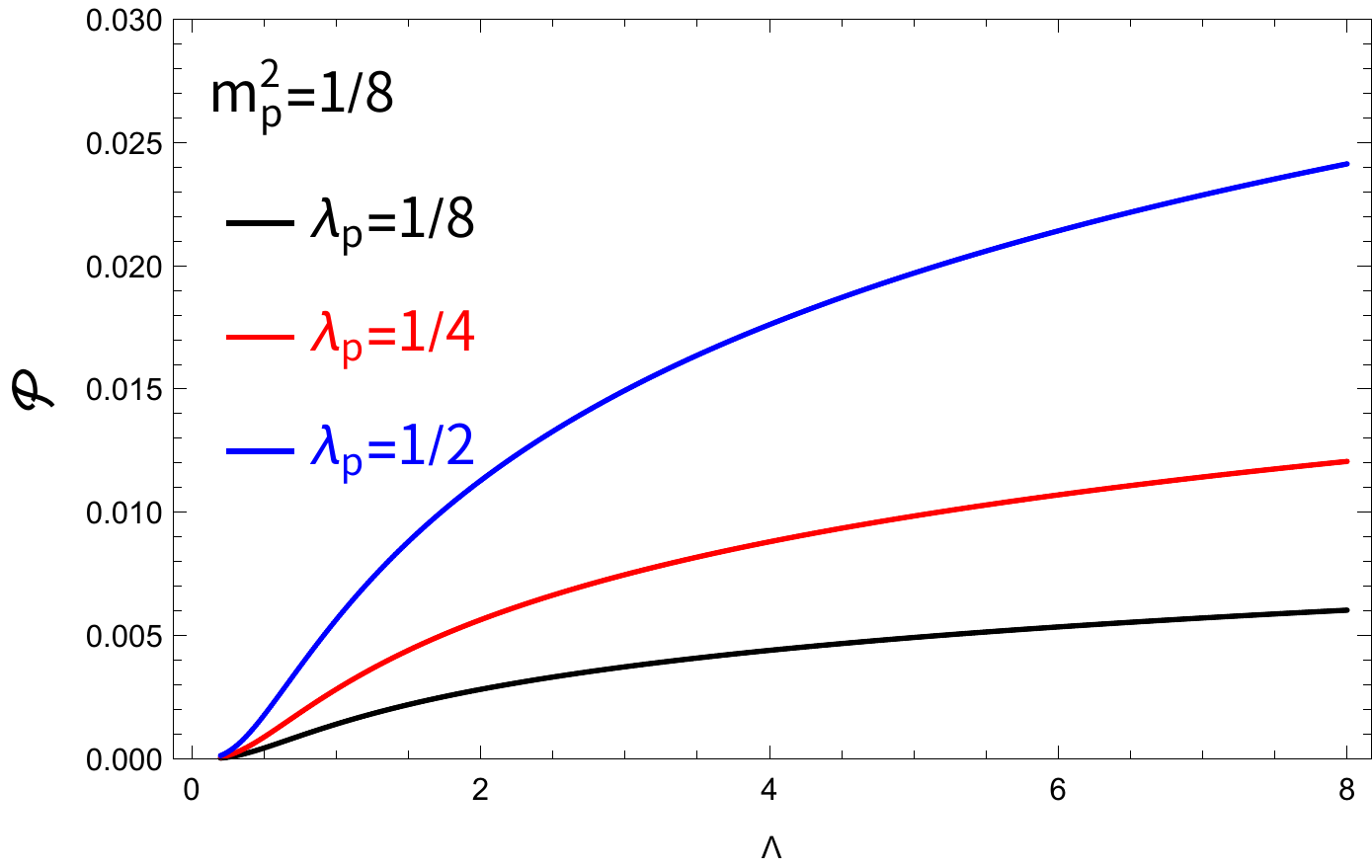}
    \includegraphics[width=0.49\textwidth]{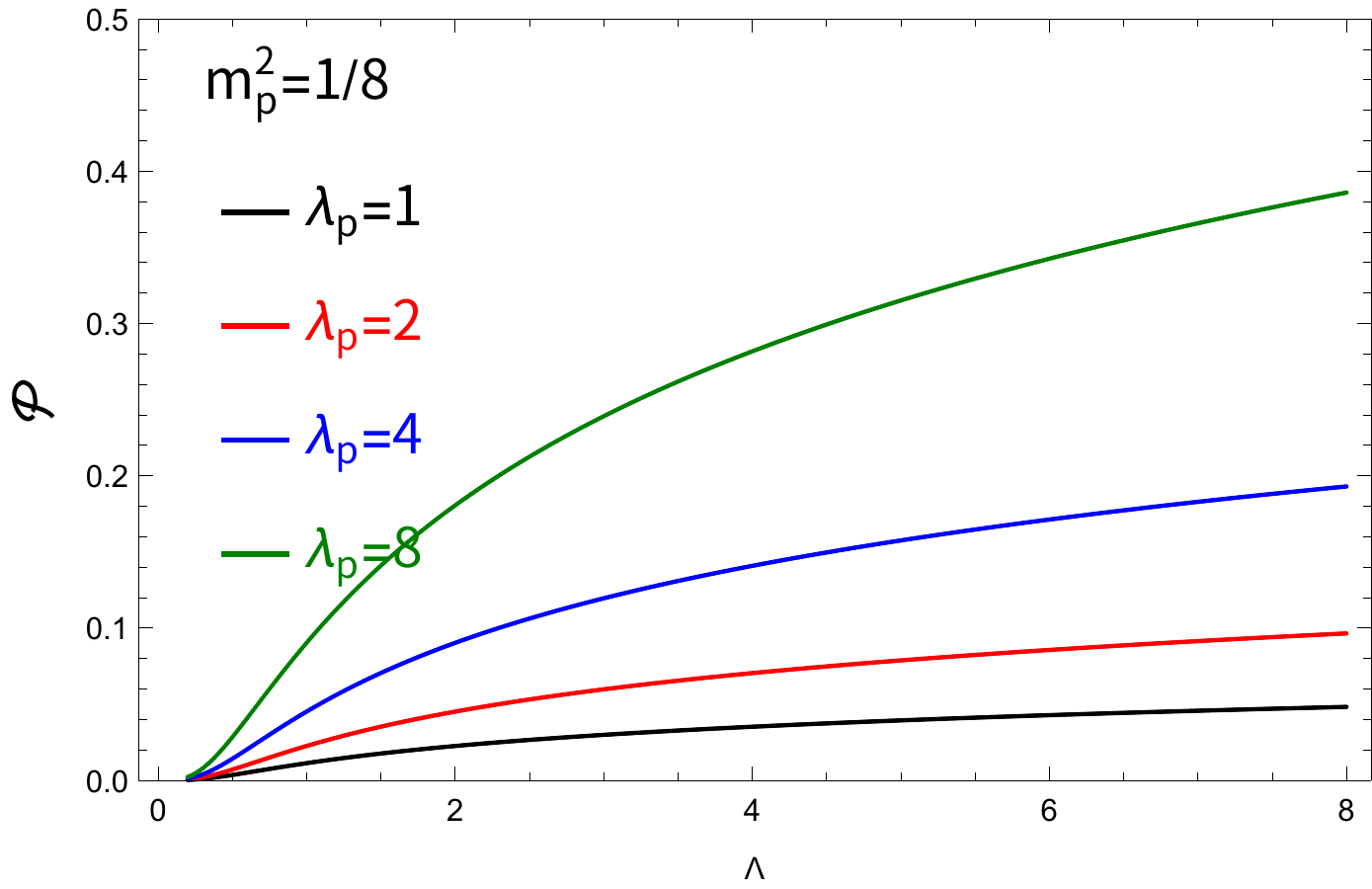}
\end{center}
\caption{We show the bare parameters and $\mathcal{P}$ for $\Lambda$. 
Because $\gamma^2(0)$ is always positive, the SSB never occurs from our current loop analysis. 
The $\lambda$ has a monotonic increasing behavior for $\Lambda$. 
In the plot of $m^2$, the solid line is always above the dashed line. 
This result implies that the RG flow never reaches the $\gamma^2(0) = 0$.}
\label{rgf}
\end{figure}

\noindent
In the situation with $\phi_0 = 0$, we have the physical mass given by
\bea
 &&
 m_p^2
 \nn\\
 &=&\gamma^2(0)
 \nn\\
 &&-
 \frac{\lambda_p^2}{3!}\frac{1}{(16\pi^2)^2}
 \nn\\
 &&\times
 \int_0^1 dx_1\int_0^{1-x_1}dx_2\ \bigg\lbrack \frac{\Lambda^4}{\alpha_1\alpha_2\big((1-\alpha_3)m_p^2+(\alpha_1+\alpha_2)\Lambda^2\big)}
  \nn\\
  &&
 +\frac{(1-\alpha_3)m_p^2}{\alpha_1^2\alpha_2^2}
 \nn\\
 &&\times
 \ln\bigg(\frac{(1-\alpha_3)m_p^2\big((1-\alpha_3)m_p^2+(\alpha_1+\alpha_2)\Lambda^2\big)}{\big((1-\alpha_3)m_p^2+\alpha_1\Lambda^2\big)
 \big((1-\alpha_3)m_p^2+\alpha_2\Lambda^2\big)
 }\bigg)\bigg\rbrack, 
 \nn\\
 \label{pss}
\eea
where 
\bea
&&
\alpha_1=x_1+x_3, \ 
\nn\\
&&
\alpha_2=\frac{x_1x_2+x_2x_3+x_3x_1}{x_1+x_3}, \ 
\nn\\
&&
\alpha_3=\frac{x_1x_2x_3}{x_1x_2+x_2x_3+x_3x_1}, 
\nn\\ 
&& 
x_1+x_2+x_3=1. 
\eea 
The $\Lambda$ is a momentum cut-off, and $\lambda_p$ is the physical coupling constant. 
We obtain the expression by identifying the physical parameters as the renormalized parameters at $\Lambda=0$. 
This equation shows that when $m_p^2 \ge 0$, $\gamma^2(0)$ must be positive since the second term is negative, inferring from the numerical computation. 
The boundary between the two regions satisfies the continuation version of Eq. \eqref{cp}. 
The solution of RG flow \cite{Gell-Mann:1954yli} (Fig. \ref{rgf}) always shows the trivial condensation.
Therefore, we only consider the $\phi_0=0$ case here.
As a result, there is no spontaneous symmetry breaking since $\langle \phi\rangle \propto \phi_0$.
\\
 
 \noindent
Next, we turn to the physical coupling $\lambda_p$, determined by computing the connected four-point function. 
Performing the integration up to two-loop, we show the result in Fig. \ref{rgf} where we illustrate the relation between $\lambda_p$ and the parameters.
To estimate the Landau pole (when $\mathcal{P}=1$), we use a quantity $\mathcal{P}$ that involves finding the absolute value of the ratio between one-loop and tree results.   
\\

\noindent 
We summarize the dependence of $\gamma^2(0)$, $\lambda$, $m^2$, and the $\mathcal{P}$ on the cutoff $\Lambda$ in Fig. \ref{rgf}. 
$\gamma^2(0), \lambda$, and ${\cal P}$ monotonically increase for $\Lambda$. 
The bare coupling constant $\lambda$ is monotone for $\Lambda$, signaling quantum triviality.
The transition line in Fig. \ref{Transition} corresponds to $\gamma^2(0)=0$. 
When substituting this condition into Eq. \eqref{pss}, the physical requirement $m_p^2 \ge 0$ forces $\lambda_p=0$ since the coefficient multiplying $\lambda_p^2$ is negative. 
The quantum triviality happens, protected by the transition line $\gamma^2(0) = 0$ (Gaussian fixed-point).
\\

\noindent
The above analysis shows that the VEV $\langle \phi\rangle$ vanishes.
Therefore, there is no spontaneous symmetry breaking in the $\phi^4$ theory alone, even with quantum corrections included.
Hence we expect that the symmetry breaking, or said differently, a nonzero $\langle \phi\rangle$ in our nature, should arise from the interaction between the Higgs field with other matter fields.

\section{Outlook}
\label{sec:6}
\noindent 
We provided a method for quantizing and analyzing strongly coupled field theories using an adaptive perturbation approach \cite{Weinstein:2005kw,Weinstein:2005kx} and diagrammatic methods. 
This method likely involves perturbing the theory away from known solvable points and treating the perturbation as a small parameter to perform calculations. 
Instead of quantizing scalar fields directly, we quantized the fields that act on states. 
This approach simplifies the calculations and makes them more tractable. 
The quantization method is extended to handle interacting theories. 
This is crucial as most real-world field theories involve interactions between fields. 
Instead of organizing the computations based on operators, we devised diagrammatic methods. 
Diagrams can be a powerful tool to visualize and resum certain types of calculations, making them more manageable. 
The diagrammatic technique involves resumming one- and two-point functions. 
The resummation likely helps capture strongly coupled effects and enhance the accuracy of the calculations. 
The results obtained using our method compared to lattice simulations. 
The fact that they matched provides evidence for the validity and efficacy of our approach.  
The resummation technique can be easily applied to fermion field theories, suggesting that our approach is versatile and can be used for a broad range of field theories. 
Based on the success in handling various aspects of field theories and matching lattice simulations, we conclude that our approach has generic applicability in strongly coupled field theories. 
This implies that it can be utilized to study and understand a wide class of systems where standard perturbative methods might not be sufficient. 
Overall, our work appears to be a promising and powerful approach to tackling strongly coupled field theories, allowing for more accurate and efficient calculations. 
It would be interesting to see how our method can be further applied and tested in different scenarios and extended to other field theories.
\\ 

\noindent 
Our computation method provides evidence of the loss of SSB in $\lambda\phi^4$ theory.
The self-interaction term is not enough to generate a non-trivial vacuum.
Experimental observation is consistent with the SSB.
Therefore, our results suggest that the generation of SSB needs other matters interaction.
  
\section*{Acknowledgments}
\noindent 
We thank Su-Kuan Chu, Bo Feng, Chang-Tse Hsieh, Xing Huang, and Chen Zhang for their helpful discussion. 
Chen-Te Ma would like to thank Nan-Peng Ma for his encouragement. 
CTM acknowledges the DOE grant (Grant No. DE-SC0021892); 
YST Program of the APCTP; 
Post-Doctoral International Exchange Program (Grant No. YJ20180087);  
China Postdoctoral Science Foundation, Postdoctoral General Funding: Second Class (Grant No. 2019M652926); 
Foreign Young Talents Program (Grant No. QN20200230017). 
YP acknowledges the National Natural Science Foundation of China (Grant No. 11905301); 
Fundamental Research Funds for the Central Universities, Sun Yat-Sen University (Grant No. 2021qntd27). 
HZ acknowledges the Guangdong Major Project of Basic and Applied Basic Research (Grant No. 2020B0301030008), 
the Science and Technology Program of Guangzhou (Grant No. 2019050001), 
and the National Natural Science Foundation of China (Grant Nos. 12047523 and 12105107). 
Discussion during the workshop ``The 15th Asia Pacific Physics Conference (APPC15)'' was helpful for this work.



\baselineskip 22pt
\begin{thebibliography}{99}
\bibitem{Goldstone:1961eq}
J.~Goldstone,
``Field Theories with Superconductor Solutions,''
Nuovo Cim. \textbf{19}, 154-164 (1961)
doi:10.1007/BF02812722

\bibitem{Luscher:1987ay}
M.~Luscher and P.~Weisz,
``Scaling Laws and Triviality Bounds in the Lattice phi**4 Theory. 1. One Component Model in the Symmetric Phase,''
Nucl. Phys. B \textbf{290}, 25-60 (1987)
doi:10.1016/0550-3213(87)90177-5

\bibitem{Coleman:1973jx}
S.~R.~Coleman and E.~J.~Weinberg,
``Radiative Corrections as the Origin of Spontaneous Symmetry Breaking,''
Phys. Rev. D \textbf{7}, 1888-1910 (1973)
doi:10.1103/PhysRevD.7.1888

\bibitem{Aizenman:1981du}
M.~Aizenman,
``Proof of the Triviality of phi**4 in D-Dimensions Field Theory and Some Mean Field Features of Ising Models for D\ensuremath{>}4,''
Phys. Rev. Lett. \textbf{47}, 1-4 (1981)
doi:10.1103/PhysRevLett.47.1

\bibitem{Weinstein:2005kw}
M.~Weinstein,
``Adaptive perturbation theory. I. Quantum mechanics,''
[arXiv:hep-th/0510159 [hep-th]].

\bibitem{Weinstein:2005kx}
M.~Weinstein,
``Adaptive perturbation theory: Quantum mechanics and field theory,''
Nucl. Phys. B Proc. Suppl. \textbf{161}, 238-247 (2006)
doi:10.1016/j.nuclphysbps.2006.08.059
[arXiv:hep-th/0510160 [hep-th]].

\bibitem{Ma:2019pxd}
C.~T.~Ma,
``Adaptive Perturbation Method in Quantum Mechanics,''
IOP SciNotes \textbf{2}, no.3, 035202 (2021)
doi:10.1088/2633-1357/ac12ba
[arXiv:1911.08211 [quant-ph]].

\bibitem{Ma:2020ipi}
C.~T.~Ma,
``Second-Order Perturbation in Adaptive Perturbation Method,''
JHAP \textbf{2}, no.2, 37 (2022)
[arXiv:2004.00842 [hep-th]].

\bibitem{Ma:2020syr}
C.~T.~Ma,
``Accurate Study from Adaptive Perturbation Method,''
Int. J. Mod. Phys. A \textbf{36}, no.04, 2150029 (2021)
doi:10.1142/S0217751X21500299
[arXiv:2007.09080 [hep-th]].

\bibitem{Schwinger:1951ex}
J.~S.~Schwinger,
``On the Green's functions of quantized fields. 1.,''
Proc. Nat. Acad. Sci. \textbf{37}, 452-455 (1951)
doi:10.1073/pnas.37.7.452

\bibitem{Gell-Mann:1954yli}
M.~Gell-Mann and F.~E.~Low,
``Quantum electrodynamics at small distances,''
Phys. Rev. \textbf{95}, 1300-1312 (1954)
doi:10.1103/PhysRev.95.1300
                                        
\end{thebibliography}
\end{document}